\documentclass[aps,superscriptaddress, showpacs,preprintnumbers, superscriptaddress, nofootinbibt, twocolumn]{revtex4-1}
\usepackage{graphicx}
\usepackage{dcolumn}
\usepackage{bm}
\usepackage{caption}
\usepackage{subcaption}
\usepackage{epstopdf}
\usepackage{amsmath}
\usepackage{mathtools}
\usepackage{xcolor}
\usepackage{epsfig}
\usepackage{wasysym}
\usepackage{amssymb}
\usepackage{slashed}
\usepackage{physics}
\usepackage{color}
\usepackage{booktabs}
\usepackage{array}
\usepackage{tabularx}
\usepackage{multirow}
\usepackage{ragged2e}
\usepackage{braket,booktabs}
\usepackage{braket,amsmath}
\usepackage{float}
\usepackage[noabbrev]{cleveref}
\usepackage{lipsum}

\begin{document}


\title{Scalar Instabilities Inside The Extremal Dyonic Kerr-Sen Black Hole: \\Novel Exact Solutions and Chronology Protection Conjecture}

\author{David Senjaya}
\email{davidsenjaya@protonmail.com}
\affiliation{High Energy Physics Theory Group, Department of Physics,
Faculty of Science, Chulalongkorn University, Bangkok 10330, Thailand}
\author{Teephatai Bunyaratavej}
\email{teephatai@gmail.com}
\affiliation{High Energy Physics Theory Group, Department of Physics,
Faculty of Science, Chulalongkorn University, Bangkok 10330, Thailand}
\author{Piyabut Burikham}
\email{piyabut@gmail.com}
\affiliation{High Energy Physics Theory Group, Department of Physics,
Faculty of Science, Chulalongkorn University, Bangkok 10330, Thailand}


\date{\today}

\begin{abstract}
 We investigate the stability of test scalar fields in the region inside the extremal Dyonic Kerr-Sen black hole (DKSBH) horizon, where closed timelike curves exist. We successfully find and present the novel exact solutions to the Klein-Gordon equation in the extremal DKSBH spacetime in terms of the Double Confluent Heun functions. The spacetime stability is explored by investigating the scalar's quasiresonance~(QS) frequencies obtained from polynomial condition of the Double Confluent Heun function. We found that both massive and massless scalar quasiresonances are double branched, having purely positive and negative imaginary frequencies, therefore, do not propagate, prohibiting time travel and suggesting no violation of Hawking's Chronology Protection Conjecture (CPC). However, only the positive branch with $0\leq\Omega_0<2(n+1)$ and negative branch with $\Omega_0>2(n+1)$ that grow exponentially has the ability to destroy spacetime. Remarkably, a new mass scale $M_{p}^{2}/M$, where $M$ is the black hole mass, is found to play a crucial role. The QS zeroth modes flip sign between purely damping and purely growing when the scalar mass is at this mass scale. 

\end{abstract}

\maketitle

\section{Introduction}
\label{sec:intro}
Stephen Hawking introduced his Chronology Protection Conjecture (CPC) in 1992 after conducting a semi-classical investigation into the stability of traversable wormholes \cite{Hawking}. Hawking's analysis of a radiation beam entering one mouth of a wormhole, while accounting for vacuum fluctuations, finds that the beam would automatically realign itself before reaching the other end of the wormhole. This phenomenon implies that the concentration of radiation becomes big enough to cause the wormhole to collapse.  According to Hawking's conjecture, there will be a physical process that can prevent the creation of a time machine, and consequently, perturbations will cause instability in the area where the time machine is located. Time travel will be prevented by the distortion of spacetime caused by the gravitational back reaction, which will blueshift the fields entering the previously described region~\cite{Vis92}. The destruction of the spacetime containing closed-timelike curves~(CTCs) makes the universe safe for historians \cite{Woodward}.

Quite recently, it is also discovered that relativistic bosonic perturbation to a wormhole throat causes bifurcation, resulting in either an inflationary universe or a black hole, depending on the total input energy, negative or positive  \cite{Shinkai,Novikov}. Nonetheless, the CPC remains inconclusive because the investigation is done in semi-classical quantum gravity, which is not reliable in high frequency regime and brief time intervals close to the Planck scale \cite{Ashok,Boris}. Full quantum gravity theory is required for ultimate conclusion. 

In the previous work \cite{Bunyaratavej:2024qgk}, we did comprehensive investigation to the most general axisymmetric black hole solution of the string inspired Einstein-Maxwell-dilaton-axion (EMDA) theory of gravity named the DKSBH. We discover that CTCs are inherent in the region behind the Cauchy horizon of rotating black holes. We did a comprehensive exploration to the sub-extremal DKSBH by carrying out a fully relativistic investigation into the stability of scalar field dynamics in the area of the black hole inner horizon, where CTCs exist. We succesfully obtained exact solutions to the Klein-Gordon equation in terms of Confluent Heun function. The exact solution allows us to find the exact quantization formula to investigate the scalar quasiresonance frequencies. However, in the extremal limit, the exact solutions break down, necessitating special attention for this particular case. The failure of numerical methods to solve the Klein-Gordon equation in extremal black hole spacetimes is also a well-known problem \cite{Richartz,Joykutty} that compels the numerical investigation to stop short only at the near-extremal limit \cite{NE1,NE2,NE3,Ponglertsakul:2020ufm}. 

In this work, we will focus on the extremal DKSBH. It is worth noting that this study presents and investigates, for the first time, the exact solutions to the Klein-Gordon equation behind the horizon of the extremal DKSBH. We will begin with a concise discussion on the closed timelike curve in the extremal DKSBH followed by rigorous derivation to get the exact solutions of the Klein-Gordon equation will be presented in detail. We firstly apply the separation of variables ansatz and successfully express the solution of the temporal part in terms of harmonic function also the angular part in terms of spheroidal harmonics. After a straightforward algebra, we successfully obtain the exact solutions to the radial wave equation in terms of the Double Confluent Heun functions. We find purely positive and negative imaginary quasiresonance frequencies for both massive and massless scalars that prohibit time travel as shown in Fig.~\ref{fig1} and \ref{fig2}, in accordance with Hawking's Chronology Protection Conjecture (CPC). Certain QS modes also grow exponentially in time and has potential to backreact and destroy the spacetime region with the CTCs.

\section{The Dyonic Kerr-Sen Black Hole}
The DKSBH is the most general axisymmetric black hole solution of the low energy effective heterotic string action in four dimensions,  which extends the Einstein-Maxwell theory by introducing a pseudo-scalar axion field coupled to the dilaton field and a coupling between the Maxwell electromagnetic field tensor and the scalar dilaton field. 

The following is the line element in Boyer-Lindquist coordinates that describes the rotating DKSBH spacetime \cite{Wu,Jana},
\begin{multline}
ds^2=-\left[1-\frac{r_s\left(r-d\right)-r^2_D}{\rho^2}\right]c^2dt^2\\
-2\frac{r_s\left(r-d\right)-r^2_D}{\rho^2}a{{\sin }^2 \theta\ }d\phi cdt+\frac{\rho^2}{\Delta }dr^2+\rho^2d\theta^2\\+\left[r\left(r-2d\right)-k^2+a^2+\frac{r_s\left(r-d\right)-r^2_D}{\rho^2}a^2{{\sin }^2 \theta\ }\right]\times \\{{\sin }^2 \theta\ } d\phi^2, \label{metric}
\end{multline}
where,
\begin{gather}
\rho^2=r\left(r-2d\right)-k^2+a^2{{\cos }^2 \theta\ },\\
\Delta =r\left(r-2d\right)-r_s\left(r-d\right)-k^2+a^2+r^2_D. \label{Delta}
\end{gather}

The black hole mass, spin/angular momentum per unit black hole mass, electric, magnetic/dyonic, dilaton, and axion charge are denoted by the variables $M, J, Q, P, d$ and $k$ and we have introduced the following notations,
\begin{gather}
r_s=\frac{2GM}{c^2},\\
r^2_D=Q^2+P^2,\\
k=2\frac{PQ}{r_s},\\
d=\frac{P^2-Q^2}{r_s},\\
a=\frac{J}{M}.
\end{gather}
Note that similar to the Kerr BH, we can analytically continue the $r$ coordinate to $r<0$ since $\rho^{2},\Delta >0$ and $g_{tt}<0$ for such negative $r$ region. The spacetime metric becomes asymptotically flat with no horizons. 

We will use $r_+$ and $r_-$ as the locations of the black hole event and Cauchy horizons, respectively, where $r_\pm$ are exactly the roots of $\Delta=0$,
\begin{gather}
r_\pm=\frac{r_s}{2}+\frac{P^2-Q^2}{r_s}\pm\frac{1}{2}\delta_r,\\
\delta_r=\sqrt{\frac{\left(r_s^2-2\left(P^2+Q^2\right)\right)^2}{r_s^2}-4a^2}. \label{dr}
\end{gather}

When $\delta_r=0$, the extreme form of the DKSBH appears, causing both of its horizons to converge at $r_H$,
\begin{equation}
  r_H=\frac{r_s}{2}+\frac{P^2-Q^2}{r_s}. \label{horizonmerged}
\end{equation}

The following equivalent alternatives can be used to express the condition $\delta_r=0$,
\begin{gather}
{\left(\frac{r_s}{2}\right)}^2+d^2+k^2-a^2-r^2_D=0, \\
{\left(r_s-\frac{2\left(P^2+Q^2\right)}{r_s}\right)}^2-4a^2=0, \label{new}
\end{gather} 
or, algebraically solving equation \eqref{new} for $r_s>0$, we obtain,
\begin{equation}
    r_s = \pm a + \sqrt{a^2+2 \left(P^2 + Q^2\right)}.  \label{coneqn}
\end{equation}

These two mass branches of extremal DKSBH were first found in Ref.~\cite{Sakti:2022izj}.

Following some lines of algebras, the metric determinant $g=\operatorname{det}(g_{\mu\nu})$ and the metric inverse $g^{\mu\nu}$ are obtained as follows \cite{Senjaya},
\begin{equation}
     g=-\rho^4{{\sin }^2 \theta\ },
\end{equation}
and,
\begin{gather}
g^{\mu\nu}=\left( \begin{array}{cccc}
-\frac{f^{\phi\phi}}{\Delta } & 0 & 0 & \frac{g_{0\phi}}{\Delta \sin^2 \theta} \\
0 & \frac{\Delta }{\rho^2} & 0 & 0 \\
0 & 0 & \frac{1}{\rho^2} & 0 \\
\frac{g_{0\phi}}{\Delta \sin^2 \theta} & 0 & 0 & {\frac{1}{\Delta {\sin }^2\theta} \left(1-\frac{r_s\left(r-d\right)-r^2_D}{\rho^2}\right)\ } \end{array}
\right), \label{metricinverse}\\
f^{\phi\phi}=r\left(r-2d\right)-k^2+a^2+\frac{r_s\left(r-d\right)-r^2_D}{\rho^2}a^2{{\sin }^2 \theta\ }.
\end{gather}

\section{Closed Timelike Curve}
Closed Timelike Curve (CTC) is a closed curve in four-dimensional spacetime with timelike tangent vectors at all points and inherent in rotating spacetimes, such as Kerr, Kerr-Newman, Kerr-Sen and their dyonic versions~\cite{Chandrasekhar}. For such axial symmetric metric, the condition for CTC to exist is mathematically described by,
\begin{equation}
    g_{\phi\phi} < 0.
\end{equation}

In our previous work \cite{Bunyaratavej:2024qgk}, we discovered that the sub-extremal DKSBH possesses either 2 or 4 CTC boundaries on the equatorial plane, depending on its charges. Similar to the sub-extremal DKSBH, the extremal DKSBH has two CTC boundaries if the black hole is lightly charged and four if it is heavily charged. The significant distinction of the extreme DKSBH is that moderate charge configuration does not make the extremal DKSBH horizon vanish. 

Figure~\ref{fig3}, \ref{fig4} and \ref{fig5} show the behavior of the metric components $g_{rr}$ and $g_{\phi\phi}$ of the two possible extremal black hole configurations. For the small charge configuration, the horizon is at $r=1.00$, while the CTC boundaries are at $r=-1.0, -0.01$. In the moderate charge configuration, the horizon is at $r=1.00$, while the CTC boundarites are at $r=-1.0, -0.09$.
Finally, in the large charge configuration, the horizon is at $r=2.045$, while the CTC boundaries are at $r=-0.332, -0.160, 0.045, 2.217$. This demonstrates that CTCs always exist in the region behind the extreme DKSBH horizon for all configurations.

\subsection{Causal Structure of DKS Spacetime}

As mentioned in Ref.~\cite{Bunyaratavej:2024qgk}, there are two singularities in DKS spacetime at $\rho^{2} =0$,
\begin{equation}
    r_{pm}=d\pm \sqrt{d^{2}+k^{2}-a^{2}\cos^{2}\theta},
\end{equation}
which are ellipsoid timelike 2-D surfaces in Boyer-Lindquist coordinates. The singularities reduce to ring-like Kerr BH singularity only when $d=k=0~(P=Q=0)$. We can define shifted DKS ellipsoidal coordinates,
\begin{eqnarray}
    x &&= \sqrt{(r-d)^{2}+a^{2}}\sin\theta\cos\phi, \notag \\
    y &&= \sqrt{(r-d)^{2}+a^{2}}\sin\theta\sin\phi,  \notag \\
    z &&= (r-d)\cos\theta,
\end{eqnarray}
where the singularities are at,
\begin{equation}
    x^{2}+y^{2}+z^{2} = d^{2}+k^{2}-a^{2}\cos(2\theta).  \label{singeq}
\end{equation}
In this DKS ellipsoidal coordinates, the two singularities are located at the single $R^{2}\equiv x^{2}+y^{2}+z^{2}=d^{2}+k^{2}-a^{2}\cos(2\theta)$ double surfaces. The semi-minor axis at $z=0~(\theta=0)$ is $\sqrt{d^{2}+k^{2}-a^{2}}$ and the semi-major axis along the $z$-direction~($\theta=\pi/2$) is $\sqrt{d^{2}+k^{2}+a^{2}}$. For extremal DSKBH, we have additional relation,
\begin{equation}
    a^{2}=\Big(\frac{r_{s}}{2}-\frac{r_{D}^{2}}{r_{s}}\Big)^{2},
\end{equation}
in the above expression~\eqref{singeq}. The semi-minor and semi-major axis now become, 
\begin{eqnarray}
   R_{\rm minor}&&=\sqrt{P^{2}+Q^{2}-\left(\frac{r_{s}}{2}\right)^{2}},  \notag \\
   R_{\rm major}&&=\sqrt{\frac{2 \left(P^2+Q^2\right)^2}{r_{s}^2}-P^2-Q^2+\left(\frac{r_{s}}{2}\right)^2},\notag
\end{eqnarray}
respectively. Note that even there is one single horizon, there are still two singularities~(see e.g. Figure~\ref{fig3}-\ref{fig5}). These singularities are separating the region $r>r_{p}$ from the region $r<r_{m}$, where notably $g_{rr}$ becomes timelike in $r_{m}<r<r_{p}$. In contrast to subextremal DKSBH where $g_{rr}$ flips sign and becomes timelike when passing the outer horizon, extremal horizon does NOT change sign of $g_{rr}$. Test particle can evade the timelike outer singularity. However, the outer singularity has repulsive (anti)gravity, and all particles following timelike geodesics should be repelled away from it. Since the outer singularity is a timelike closed 2-D surface, nongeodesic timelike curves can reach the singularity where spacetime breaks down, in contrast to ring singularity of Kerr BH where the particle can bypass to the inner region via $\theta\neq \pi/2$ trajectories. Extremal DKSBH has distinctive causal structure from Kerr BH in this aspect, every particle has to pass outer singularity in order to reach inner region $r<r_{p}$. And currently we do not know what would happen at the singularity. 

For large charge extremal DKSBH, e.g. depicted in Figure~\ref{fig5}, this is the naked singularity solution. The region $r>r_{p}$ is a regular spacetime with positive $g_{rr}$. Behind the outer singularity, $g_{rr}<0$~(since $\rho^2$ flips sign) and $g_{tt}$ is negative, so $r$ and $t$ are timelike, there are now two timelike directions. Slightly more inside, there also exists CTCs where $g_{\phi\phi}<0$, i.e., there are now three timelike directions~(one is CTC) and one spacelike direction. For smaller $r$, $g_{tt}$ becomes spacelike and two timelike directions remain. This exotic region is interesting but in this work, we will focus on the region where $g_{rr}>0,g_{tt}<0$ with existence of CTCs. The region between two singularities will be explored in the future work. Note that even when there is naked singularity, the spacetime always contain the region behind the inner singularity $r<r_{m}$ where $g_{rr}>0,g_{tt}<0$ and $g_{\phi\phi} <0$ where CTC exists and our exact solutions are applicable.

\section{Relativistic Scalar Field}
To investigate the propagation of massive and massless scalar fields, $\psi$, let us begin with a discussion on the equation of motion. In this context, the scalar fields are thought to have no additional interactions except to be weakly coupled to gravity. Therefore, the dynamics of scalar field $\psi$ in a curved spacetime is described by the following covariant Klein–Gordon equation \cite{Derig,Luzio},
\begin{equation}
        \left[\nabla_\mu\nabla^\mu - \left(\frac{mc}{\hbar}\right)^2\right]\psi = 0,
    \end{equation}
where $\nabla_\mu$ is covariant derivative and $m$ is the scalar's rest mass. It is possible to explicitly express the covariant wave equation in terms of partial derivatives via the Laplace-Beltrami operator as follows,
    \begin{equation}
        \left[\frac{1}{\sqrt{-g}}\partial_\mu\left(\sqrt{-g}\ g^{\mu\nu}\partial_\nu\right) - \left(\frac{mc}{\hbar}\right)^2\right]\psi = 0. \label{KG}
    \end{equation}

Substituting the metric determinant $g={det}(g_{\mu\nu})$ and the metric inverse $g^{\mu\nu}$ into \eqref{KG}, we obtain this following expression (see \cite{Senjaya} for derivation details), \\
\begin{multline}
\left[-\frac{{1}}{\Delta \rho^2}\left\{{\left[r\left(r-2d\right)-k^2+a^2\right]}^2-\Delta a^2{{\sin }^2 \theta\ }\right\}{\partial }^2_{ct}\right. \\ \left.-2\frac{\left[r\left(r-2d\right)-k^2+a^2-\Delta \right]a}{\Delta \rho^2}{\partial }_{ct}{\partial }_\phi\right. \\ \left.+\frac{1}{\rho^2}{\partial }_r\left(\Delta {\partial }_r\right)+\frac{1}{\rho^2{\sin  \theta\ }}{\partial }_\theta\left({\sin  \theta\ }{\partial }_\theta\right)\right. \\ \left.+\frac{\Delta -a^2{{\sin }^2 \theta\ }}{\Delta {{\sin }^2 \theta\ }\rho^2}{\partial }^2_\phi\right]\psi-\frac{m^2 c^2}{{\hbar }^2}\psi=0. \label{fullwave}
\end{multline}

The expression above is a linear second order differential equation, which depends on the spacetime variables $ct,r,\theta,\phi$. Using the separation ansatz, we can express the scalar field $\psi(ct,r,\theta,\phi)$, taking into account the temporal and azimuthal symmetry \cite{Dong,35},
\begin{gather}
\psi\left(t,r,\theta,\phi\right)=e^{-i\frac{E}{\hbar c}ct+im_\ell \phi}R\left(r\right)T\left(\theta\right).
\end{gather}

Substituting back the ansatz into the main equation \eqref{fullwave}, followed by multiplying the whole equation by $\rho^2r^2/\psi \left(t,r,\theta ,\phi \right)$, we arrive at this following expression,
\begin{multline}
\left[\frac{1}{T{\sin  \theta\ }}{\partial }_\theta\left({\sin  \theta\ }{\partial }_\theta T\right)-\frac{m^2_l}{{{\sin }^2 \theta\ }}\right. \\ \left.-\left(\frac{\Omega^2_0 a^2}{r^2_s}-\frac{\Omega^2a^2}{r^2_s}\right){{\cos }^2 \theta\ }\right]+\left[\frac{1}{R}{\partial }_r\left(\Delta {\partial }_rR\right)\right. \\ \left.+\frac{\Omega^2}{r^2_s}\frac{{\left(r\left(r-2d\right)-k^2+a^2\right)}^2}{\Delta }-\frac{\Omega^2a^2}{r^2_s}\right. \\ \left.-2\frac{\left(r\left(r-2d\right)-k^2+a^2-\Delta \right)a}{\Delta }\left(\frac{\Omega m_\ell }{r_s}\right)\right. \\ \left.+\frac{m^2_l a^2}{\Delta }-\frac{\Omega^2_0}{r^2_s}\left(r\left(r-2d\right)-k^2\right)\right]=0,
\end{multline}
where we have defined two dimensionless energy parameters,
\begin{equation}
   \Omega=\frac{Er_s}{\hbar c} \ \ \ , \ \ \Omega_0=\frac{E_0r_s}{\hbar c}=\frac{(mc^2)r_s}{\hbar c}.
\end{equation}
Note that $\Omega_{0}$ can be understood as the inverse ratio of the Compton wavelength of the scalar and the Schwarzschild circumference of the black hole.

\subsection{Polar Sector}
The first squared bracket depends only on $\theta$, therefore, can be separated from the rest as follows,
\begin{multline}
  \frac{1}{{T\sin  \theta\ }}{\partial }_\theta\left({\sin  \theta\ }{\partial }_\theta T\right)-\frac{m^2_\ell}{{{\sin }^2 \theta\ }}\\-\left(\frac{\Omega^2_0a^2}{r^2_s}-\frac{\Omega^2a^2}{r^2_s}\right){{\cos }^2 \theta\ }=-\lambda^{m_\ell }_\ell ,   \label{fullwave1}
  \end{multline}
where $\lambda^{m_\ell }_\ell $ is the separation constant. 

Remark that if $a=0$, $\lambda^{m_\ell }_\ell =\ell\left(\ell+1\right)$ and the exact solution of the polar sector is given by the associated Legendre polynomial, $P^{m_\ell }_\ell (\cos \theta)$. If combined with the azimuthal solution, the whole angular solution becomes the Spherical Harmonics.

However, for a general rotating black hole spacetime, the polar equation is expressed as a superposition of the Spherical Harmonics called Spheroidal Harmonics, $S^{m_\ell }_l$,
\begin{gather}
T(\theta){=}S^{m_\ell }_\ell \left(\sigma,{\cos  \theta\ }\right)=\sum^{\infty }_{r=-\infty }{d^{\ell m_\ell }_r\left(\sigma\right)P^{m_\ell }_{\ell +r}\left({\cos \theta}\right)},
\end{gather}
where,
\begin{equation}
 \sigma= \frac{\Omega^2_0a^2}{r^2_s}-\frac{\Omega^2a^2}{r^2_s}.
\end{equation}

The coefficients $d^{lm_\ell }_r$ is a constant and act as amplitude of the corresponding mode in the expansion. For the case $\sigma<1$, $\lambda^{m_\ell }_\ell $ can be calculated perturbatively  \cite{Press,Berti1,Berti2,Cho:2009wf,Suzuki:1998vy}, and expressed in this following expression, 
\begin{equation}
\lambda^{m_\ell }_\ell =\ell(\ell+1)-2\sigma\left(\frac{m_\ell ^2+\ell(\ell+1)-1}{(2\ell-1)(2\ell+3)}\right)+O\left(\sigma^2  \right).
\end{equation}

\subsection{Radial Sector}
Replacing the polar sector in the equation \eqref{fullwave1} by the separation constant $\lambda^{m_\ell }_\ell$, we are left with the following radial equation,
\begin{multline}
{\partial }_r\left(\Delta {\partial }_rR\right)+\left[\frac{1}{\Delta }{\left\{\frac{\Omega}{r_s}\left(r\left(r-2d\right)-k^2+a^2\right)-m_\ell a\right\}}^2\right. \\ \left.-\left\{\frac{\Omega^2_0}{r^2_s}\left(r\left(r-2d\right)-k^2+a^2\right)+\frac{\Omega^2}{r^2_s}a^2-\frac{\Omega^2_0}{r^2_s}a^2\right.\right. \\ \left.\left.-2\frac{\Omega m_\ell}{r_s}a+\lambda^{m_\ell}_\ell\right\}\right]R{=0}, \label{radialmod}
\end{multline}

It is noteworthy that the aforementioned radial equation possesses a general applicability and valid for all scenarios, encompassing sub-extremal, extremal, or super-extremal DKSBHs.

Now, let us focus our invetigation on the extremal DKSBH case. In the extreme condition, where $\delta_r=0$, the black hole’s horizons are merged at a position, $r_H$, given by equation \eqref{horizonmerged}. Therefore, we can rewrite $\Delta=\Delta_E$ as follows,
\begin{equation}
    \Delta_E=(r-r_H)^2,
\end{equation}
and consequently, we get the following expression of the radial equation in the extremal rotating black hole spacetime,
\begin{widetext}
\begin{multline}
    {\partial }^2_rR+\frac{2}{r-r_H}{\partial }_rR+\left[\frac{1}{{\left(r-r_H\right)}^4}{\left\{\frac{\Omega }{r_s}\left(a^2-k^2+r\left(r-2d\right)\right)-am_{\ell }\right\}}^2\right. \\ \left.-\frac{1}{{\left(r-r_H\right)}^2}\left\{\frac{{\Omega }^2_0}{r^2_s}\left(a^2-k^2+r\left(r-2d\right)\right)+K^{m_{\ell }}_{\ell }\right\}\right]R=0,
\end{multline}
\end{widetext}
where,
\begin{equation}
K^{m_\ell }_\ell=\frac{\Omega^2}{r^2_s}a^2-\frac{\Omega^2_0}{r^2_s}a^2-2\frac{\Omega m_\ell }{r_s}a+\lambda^{m_\ell }_\ell.  \label{Klm}
\end{equation}

Be defining a new dimensionless radial variable, $\displaystyle{x=\frac{r-r_H}{r_s}}$, we obtain the following one dimensional Schr\"odinger-like equation,
\begin{gather}
    {\partial }^2_xR+\frac{2}{x}{\partial }_xR+V_{eff}\left(x\right)R=0, \label{radx}
\end{gather}

The effective potential, $V_{eff}\left(x\right)$, is obtained as follows,
\begin{equation}
V_{eff}\left(x\right)=V_0+\frac{V_1}{x}+\frac{V_2}{x^2}+\frac{V_3}{x^3}+\frac{V_4}{x^4}, \label{Veff}
\end{equation}
where the coefficient of each term of the $V_{eff}\left(x\right)$ is given as follows,
\begin{align}
{{V}}_{{4}}&=\frac{1}{r^4_s}{\left\{\Omega \left(a^2-k^2+r_H\left(r_H-2d\right)\right)-am_{\ell }r_s\right\}}^2 \nonumber\\ 
&=\begin{multlined}[t]{\left\{\Omega -\frac{am_{\ell }r_s}{a^2-k^2+r_H\left(r_H-2d\right)}\right\}}^2\times \nonumber \\ {\left(\frac{a^2-k^2+r_H\left(r_H-2d\right)}{r^2_s}\right)}^2 \end{multlined} \nonumber \\ 
&={\left\{\Omega -{\Omega }_c\right\}}^2\zeta ^2 \nonumber\\  
&={{\Delta }}^2_{{\Omega }}{\zeta }^2,
\end{align}
where we have defined three dimensionless constants,
\begin{gather}
{\Delta }_{\Omega }=\Omega -{\Omega }_c,\\
{\Omega }_c=\frac{am_{\ell }r_s}{a^2-k^2+r_H\left(r_H-2d\right)},\\
\zeta =\frac{a^2-k^2+r_H\left(r_H-2d\right)}{r^2_s}.
\end{gather}

The coefficient of $x^{-3}$ reads,
\begin{align}
   V_3&=\begin{multlined}[t]\frac{1}{r^3_s}\Omega \left(4r_H-4d\right) \times \nonumber \\\left\{\Omega \left(a^2-k^2+r_H\left(r_H-2d\right)\right)-am_{\ell }r_s\right\} \end{multlined} \nonumber\\
   &={2}{\Omega }{{\Delta }}_{{\Omega }}\zeta ,
\end{align}
where we have used the equation for extremal black hole horizon, $\displaystyle{r_H-d=\frac{r_s}{2}}$, to obtain the final expression.

The coefficient of $x^{-2}$ after substituting $K^{m_\ell }_\ell$ as in \eqref{Klm} reads as follows,
\begin{multline}
    V_2=\frac{1}{r^2_s}\left\{-r^2_s{\lambda }^{m_{\ell }}_{\ell }+{\Omega }^2_0\left(k^2-r_H\left(r_H-2d\right)\right)\right. \\ \left.+{\Omega }^2\left(a^2+4d^2-2k^2+6r_H\left(r_H-2d\right)\right)\right\},
\end{multline}
where simplifications can be made,
\begin{multline}
    a^2+4d^2-2k^2+6r_H\left(r_H-2d\right)\\=2\left(a^2-k^2+r_H\left(r_H-2d\right)\right)-a^2+4{\left(d-r_H\right)}^2  \\=2\zeta r^2_s-a^2+4{\left(-\frac{r_s}{2}\right)}^2=r^2_s\left(2\zeta +1\right)-a^2,   
\end{multline}
and,
\begin{equation}
    k^2-r_H\left(r_H-2d\right)=a^2-\zeta r^2_s.
\end{equation}
and the final expression of $V_2$ is obtained,
\begin{equation}
V_2=-\lambda^{ m_\ell}_{\ell }+\frac{{\Omega }^2}{r_s^2}\left(r^2_s\left(2\zeta +{ 1}\right) - a^2\right)+\frac{{\Omega }^2_0}{r_s^2}\left(a^2 - \zeta r^2_s\right).
\end{equation}

The last two coefficients need slight simplification and can be presented as follows,
\begin{gather}
V_1=-\frac{1}{r_s}2\left(d-r_H\right)\left(2{\Omega }^2-{\Omega }^2_0\right) = 2\Omega^2 - \Omega^2_0,\\
V_0=-\left(\Omega^2_0 - \Omega^2\right).
\end{gather}

Putting each of the simplified coefficient back into equation \eqref{Veff}, we obtain this following compact expression,
\begin{multline}
V_{eff}\left(x\right)=-\left[{\Omega }^2_0-{\Omega }^2\right]+\frac{1}{x}\left[{2}{\Omega }^2-{\Omega }^2_0\right]\\
+\frac{1}{x^2}\left[-{\lambda }^{m_{\ell }}_{\ell }+\frac{{\Omega }^2}{r_s^2}\left(r^2_s\left(2\zeta +1\right)-a^2\right)+\frac{{\Omega }^2_0}{r_s^2}\left(a^2-\zeta r^2_s\right)\right]\\+\frac{1}{x^3}\left[2\Omega {\Delta }_{\Omega }\zeta \right]+\frac{1}{x^4}\left[{\Delta }^2_{\Omega }\zeta ^2\right].    
\end{multline}

Following the Appendix \ref{AppendixA}, we should substitute $R(x)=\frac{1}{x}{\mathcal{R}}\left(x\right)$ in order to remove the first order derivative and transform the the radial equation \eqref{radx} into its normal form as follows,
\begin{equation}
{\partial }^2_x{\mathcal{R}}+V_{eff}\left(x\right){\mathcal{R}}=0. \label{normalRadeq}
\end{equation}

Remark that the normal form of the radial equation above exactly matches the normal form of the Double Confluent Heun equation \eqref{normalDCHeq}. Therefore, the radial exact solution of the Klein-Gordon equation in the extremal DKSBH \eqref{normalRadeq} can be expressed in terms of the Double Confluent Heun functions as follows,
\begin{multline}
R={e^{\frac{1}{2}\left(\epsilon x-\frac{\gamma }{x}\right)}x^{\frac{\delta }{2}-1}}\times \\ \left[A\  \operatorname{HeunD}\left(\beta ,\alpha ,\gamma ,\delta ,\epsilon ,x\right)+Be^{\frac{\gamma }{x}-\epsilon x}x^{2-\delta }   \times \right. \\ \left.\operatorname{HeunD}\left(-2+\beta +\delta ,\alpha -2\epsilon ,-\gamma ,4-\delta ,-\epsilon ,x\right)\right]. \label{Req}
\end{multline}

Now, let us calculate the explicit expressions of the Double Confluent Heun parameters as functions of the scalar and black hole parameters. We compare the two normal forms, \eqref{normalRadeq} and \eqref{normalform}.
\begin{enumerate}
\item The coefficient of $x^0$:
\begin{equation}
-\left({\Omega }^2_0-{\Omega }^2\right)=-\frac{\epsilon ^2}{4}\to\epsilon _\pm=\pm2\sqrt{{\Omega }^2_0-{\Omega }^2}.
\end{equation}

For non-zero scalar mass, to determine the correct sign, consider the regularity of the asymptotic behavior of the wave equations \eqref{normalRadeq} and \eqref{normalDCHeq} at $x\to-\infty$, as follows,
\begin{equation}
    R(x\to-\infty)\sim e^{\pm\sqrt{{\Omega }^2_0-{\Omega }^2}x}=e^{\epsilon
    _\pm x}\to 0,  \label{asineq}
\end{equation}
clearly, the correct choice is $\epsilon_+$.

\item The coefficient  $x^{-4}$:
\begin{equation}
{\Delta }^2_{\Omega }{\zeta }^2=-\frac{{\gamma }^2}{4}\to {\gamma }_\pm=\pm2i{\Delta }_{\Omega }\zeta.
\end{equation}

\item The coefficient  $x^{-3}$:
\begin{align}
2\Omega {\Delta }^{\ }_{\Omega }\zeta &={\gamma }_\pm \left(1-\frac{\delta }{2}\right),\nonumber \\
2\Omega {\Delta }^{\ }_{\Omega }\zeta &=\pm 2i{\Delta }_{\Omega }\zeta \left(1-\frac{\delta }{2}\right),\nonumber \\
\delta &=2\left(1\pm i\Omega \right). 
\end{align}

\item The coefficient  $x^{-1}$:
\begin{align}
2{\Omega }^2-{\Omega }^2_0&=\alpha -\frac{\epsilon_+\delta }{2},\nonumber\\
2{\Omega }^2-{\Omega }^2_0&=\alpha -\frac{2\sqrt{{\Omega }^2_0-{\Omega }^2}\times 2\left(1\pm i\Omega \right)}{2},\nonumber\\
\alpha &=2{\Omega }^2-{\Omega }^2_0+2\left(1 \pm i\Omega \right)\sqrt{{\Omega }^2_0-{\Omega }^2}.
\end{align}

\item The coefficient  $x^{-2}$:
\begin{gather}
    V_2=\frac{\delta}{2}-\frac{\delta^2+2\epsilon_+ \gamma_\pm}{4}-\beta,\\
\begin{multlined}\beta=\lambda^{m_\ell}_\ell+ \Omega\left\{\left(1 + \frac{a^2 - 
       r_s^2 \left(1 + 2 \zeta\right)}{r_s^2} \right)\Omega\mp i\right\} \\+ (r_s^2 \zeta-a^2) \frac{\Omega_0^2}{r_s^2}\mp 2i\zeta \Delta_\Omega \sqrt{\Omega_0^2-\Omega^2}    
    \end{multlined}.
\end{gather}
\end{enumerate}

\subsection{Quasiresonance Frequencies}
Let us now consider the Double Confluent Heun function's polynomial condition \eqref{HeunDPol}. The regularity of the radial solution at $x\to -\infty$ is ensured by the polynomial condition, which terminates the Frobenius series expansion of the Double Confluent Heun function at order $n$. Furthermore, the formula for calculating the quasiresonance frequency will be derived from the polynomial condition that connects the radial quantum number $n$ to the black hole parameters as well as the scalar mass and energy. Substituting the explicit expression of $\alpha$ and $\epsilon$ into \eqref{HeunDPol} yields,
\begin{gather}
\frac{2{\Omega }^2-{\Omega }^2_0+2\left(1\pm i\Omega \right)\sqrt{{\Omega }^2_0-{\Omega }^2}}{2\sqrt{{\Omega }^2_0-{\Omega }^2}}=-n \ \ , \ \ n=0,1,\dots\\
\frac{2{\Omega }^2-{\Omega }^2_0}{2\sqrt{{\Omega }^2_0-{\Omega }^2}}\pm i\Omega =-(n+1).   \label{QNM}
\end{gather}
This equation has lengthy analytic solutions which reduces from 3 to 1 solution when subject to the constraint from square root in the denominator. Numerically, it is found that there is only one {\it pure imaginary} root. 

In Appendix \ref{AppendixC}, we provide an {\it analytic} proof that the solution to \eqref{QNM} is purely imaginary for $\Omega_{0}<2$. For $\Omega_{0}\geq 2$, we numerically verify up to very large rest mass that all solutions are still pure imaginary, example plots are shown in Fig.~\ref{contourfig}. Also note that the sign of the term $\pm i\Omega$ in equation \eqref{QNM} depends on the choice of $\gamma_\pm$, i.e., positive (negative) sign for to $\gamma_+ (\gamma_-)$ mode. However, the sign will flip for sufficiently large $\Omega_{0}$~(e.g. zeroth mode $n=0$ flips sign at $\Omega_{0}=2$).

Now, let us set $\Omega_0=0$ to investigate the quasiresonance frequencies of massless scalar, the energy equation \eqref{QNM} is simplified as follows,
\begin{equation}
    \frac{\Omega^2}{\pm \sqrt{-\Omega^2}} \pm i \Omega=-(n+1),
\end{equation}
where the sign in front of $\sqrt{-\Omega^2}$ is determined by $\epsilon_\pm$ and the sign of $i\Omega$ is determined by $\gamma_\pm$.

The equation can be solved algebraically and only $(\epsilon_+, \gamma_\pm)$ modes have solutions, shown below,
\begin{equation}
\Omega =\pm \frac{i}{2}\left(1+n\right) \ \ , \ \ n=0,1,\dots, \label{masslessqrf}
\end{equation}
where positive (negative) sign corresponds to $\gamma_+ (\gamma_-)$ mode. These modes also satisfy Eqn.~(\ref{asineq}), i.e., the field vanishes at $x\to -\infty$. At the horizon, the asymptotic behavior of the wave~(\ref{Req}) at $x\to 0$ becomes
\begin{equation}
R(x\to 0)\sim e^{\mp i\Delta_{\Omega}\frac{\zeta}{x}}.
\end{equation}
Using tortoise coordinate defined as,
\begin{eqnarray}
r^{*}&&=\int\frac{r(r-2d)+a^{2}-k^{2}}{(r-r_{H})^{2}}dr, \notag \\ 
r^{*}(r_{H})&&\approx \frac{r_{H}(r_{H}-2d)+a^{2}-k^{2}}{-(r-r_{H})}=\frac{\zeta}{-x}.
\end{eqnarray}
The near-horizon solution then can be expressed as follows,
\begin{equation}
R(x\to 0)\sim e^{\pm i\Delta_{\Omega} r^{*}},   
\end{equation}
corresponding to outgoing and incoming solution with respect to the horizon respectively. Namely, the $\gamma_{+}~(\gamma_{-}$ modes are the outgoing~(incoming) solution to~(from) $x\to -\infty$. With respect to the interior region of the black hole, the quasinormal modes~(QNMs) are the $\epsilon_{+},\gamma_{+}$ which exponential grows in time while the $\epsilon_{+},\gamma_{-}$ modes are the pure damping modes. The QNMs even having zero real part will enhance and backreact on the interior spacetime behind horizon of the black hole and deform it. The spacetime interior region with the presence of CTCs is unstable with respect to the the scalar perturbation. 

\begin{figure}[H]
    \centering
    \includegraphics[scale=0.6]{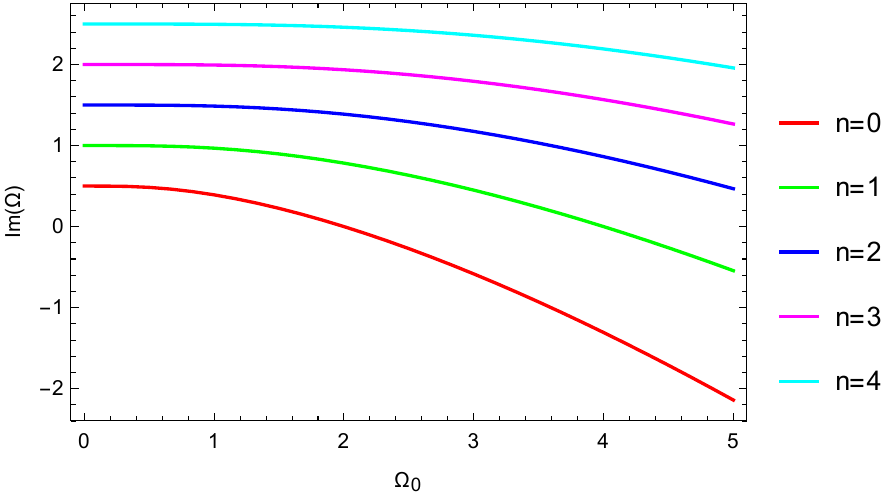}
    \caption{Quasiresonance frequencies for the first five $\epsilon_+, \gamma_+$ modes with $n=0, 1,2,3,4$. The unit of $\Omega_{0},\Omega$ is $r_{s}/\hbar c$ where $r_{s}=2GM/c^{2}$. By Eqn.~\eqref{coneqn}, each point on the curve thus corresponds to set of $(a,P,Q)$ values satisfying this extremal condition.}  
  \label{fig1}
\end{figure}

\begin{figure}[H]
    \centering
    \includegraphics[scale=0.6]{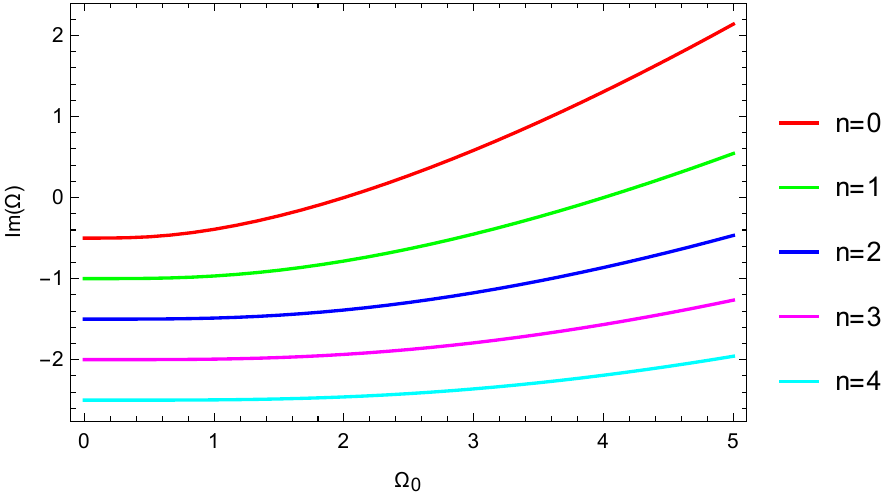}
    \caption{Quasiresonance frequencies for the first five $\epsilon_+, \gamma_-$ modes with $n=0, 1,2,3 ,4$. The unit of $\Omega_{0},\Omega$ is $r_{s}/\hbar c$ where $r_{s}=2GM/c^{2}$. By Eqn.~\eqref{coneqn}, each point on the curve thus corresponds to set of $(a,P,Q)$ values satisfying this extremal condition.}
  \label{fig2}
\end{figure}

In Figure~\ref{fig1} and Figure~\ref{fig2}, we plot the quasiresonance frequencies for the first five $\gamma_+$ and $\gamma_-$ modes respectively with $n=0-4$ against the scalar mass $\Omega_0$. All modes are pure imaginary. Notably the zeroth mode $n=0$ changes sign at $\Omega_{0}=2$. Remark that this corresponds to the scalar mass, 
\begin{equation}
m = \frac{E_{0}}{c^{2}}=\frac{\Omega_{0}\hbar c}{r_{s}c^{2}}=\frac{M_{p}^{2}}{M},
\end{equation}    
for the Planck mass $M_{p}\equiv \sqrt{\hbar c/G}$. For $M>M_{p}$, we have a hierarchy of mass scale $M_{p}^{2}/M<M_{p}<M$ and the masses are geometrically related. For a solar-mass BH, this mass is $M_{p}^{2}/M = 1.34\times 10^{-10}$ eV$/c^2$.  

The fundamental mode, $n=0$, is the most sensitive to the scalar mass. The higher the excitation, the less sensitive the mode is to the scalar mass. And for $\Omega_0\to 0$, we can easily verify the exact formula \eqref{masslessqrf} from the plot. 

The $(\epsilon_+, \gamma_+)$ modes have purely positive imaginary quasiresonance frequencies in the range $0\leq\Omega_0<2$, they grow exponentially, and can destroy the spacetime containing CTCs and decay otherwise. On the other hand, the $\epsilon_+, \gamma_-$ modes have negative imaginary quasiresonance frequencies in the range $0\leq\Omega_0<2$ that decay exponentially and grow otherwise. Both modes have zero $Re(\Omega)$, therefore, do not propagate, making it impossible to do time travel. Remark that the positive branch with $0\leq\Omega_0<2(n+1)$ and negative branch with $\Omega_0>2(n+1)$ has the $n$-th modes $\Omega_{n}$ that grow exponentially and capable of destroying the inner region of spacetime where CTC exists. 

\subsection{The Stress-Energy Tensor}
In this section, we will compute the stress-energy tensor for the Klein-Gordon field in the interior of extremal DKS spacetime where CTC exists. The stress-energy tensor of the complex scalar field is given as follows (see e.g. \cite{Baumgarte:2010ndz}),
\begin{equation}
    T_{\mu\nu}=\partial_\mu \psi \partial_\nu \psi^* + \partial_\nu \psi\partial_\mu \psi^* - g_{\mu\nu} \left(\partial_\sigma \psi^* \partial^\sigma \psi + \frac{\Omega_0^2}{r_s^2}\psi^* \psi \right).
\end{equation}

First, consider $T_{00}$. Since $\partial_0 \psi= -i \frac{\Omega}{r_s} \psi$, we obtain,
\begin{gather}
    T_{00}=\left\{2\left|\frac{\Omega}{r_s}\right|^2-g_{00}\frac{\Omega_0^2}{r_s^2}\right\}|\psi|^2 - g_{00} \partial_\sigma \psi^* \partial^\sigma \psi. \label{T00}
\end{gather}

Let us consider the last term of equation \eqref{T00},
\begin{equation}
\partial_\sigma \psi^* \partial^\sigma \psi=g^{\sigma\alpha}\partial_\sigma \psi^* \partial_\alpha \psi, \label{new}
\end{equation}
where the wave function and its inverse are given as the following,
\begin{gather}
\psi\left(t,r,\theta,\phi\right)=e^{-i\frac{\Omega}{r_s}ct}e^{im_\ell \phi}R\left(r\right)T\left(\theta\right),\\
\psi^*\left(t,r,\theta,\phi\right)=e^{i\frac{\Omega^*}{r_s}ct}e^{-im_\ell \phi}R^*\left(r\right)T^*\left(\theta\right),
\end{gather}
also note that $\Omega$ is complex valued yielding the following expressions,
\begin{align}
    i\Omega&=i\operatorname{Re}(\Omega)- \operatorname{Im}(\Omega),\\
    \left(i\Omega\right)^*=-i\Omega^*&=-i\operatorname{Re}(\Omega)- \operatorname{Im}(\Omega).
\end{align}
To find the explicit expression of the equation \eqref{new}, we can substitute explicitly the metric inverse expression \eqref{metricinverse}. It is straightforward to obtain,
\begin{multline}
\partial_\sigma \psi^* \partial^\sigma \psi=\left\{-\left|\frac{\Omega}{r_s}\right|^2\frac{f^{\phi\phi}}{\Delta}-2 m_\ell \frac{\rm Re(\Omega)}{r_s}\frac{g_{0\phi}}{\Delta \sin^2 \theta}+ \right. \\ \left.  \frac{m_\ell^2}{\Delta {\sin }^2\theta} \left(1-\frac{r_s\left(r-d\right)-r^2_D}{\rho^2}\right)   \right\} |\psi|^2 \\+\left(\frac{\Delta}{\rho^2}|T|^2|\partial_rR|^2+\frac{1}{\rho^2}|\partial_\theta T|^2|R|^2\right) e^{2\frac{\rm Im(\Omega)}{r_s}ct} . 
\end{multline}

Substituting the above expression back into \eqref{T00} along with,
\begin{equation}
 |\psi|^2 = e^{2\frac{\rm Im(\Omega)}{r_s}ct}|T|^2|R|^2 , \label{grofield}
\end{equation}
yields,
\begin{widetext}
    \begin{multline}
  T_{00}=e^{2\frac{\rm Im(\Omega)}{r_s}ct}\left[\left\{2\left|\frac{\Omega}{r_s}\right|^2+\left(1-\frac{r_s\left(r-d\right)-r^2_D}{\rho^2}\right)\left(\frac{m_\ell^2}{\Delta {\sin }^2\theta} \left(1-\frac{r_s\left(r-d\right)-r^2_D}{\rho^2}\right)+\frac{\Omega_0^2}{r_s^2}-\left|\frac{\Omega}{r_s}\right|^2\frac{f^{\phi\phi}}{\Delta} \right.\right.\right. \\ \left.\left.\left. -2 m_\ell \frac{\rm Re(\Omega)}{r_s}\frac{g_{0\phi}}{\Delta \sin^2 \theta}  \right)\right\}|T|^2|R|^2+\left(1-\frac{r_s\left(r-d\right)-r^2_D}{\rho^2}\right)\left(\frac{\Delta}{\rho^2}|T|^2|\partial_rR|^2+\frac{1}{\rho^2}|\partial_\theta T|^2|R|^2\right)\right]  .       
    \end{multline}
\end{widetext}

We can show that the quantity inside the square brackets of $T_{00}$ is generally non-zero by considering the simplest case, i.e., the $m_\ell=0$ state on the equatorial plane. By setting $\theta=\frac{\pi}{2}$ and writing $f^{\phi\phi}$, $\rho$, and $\Delta$ as functions of black hole parameters and coordinates yields the following expression,
\begin{widetext}
    \begin{multline}
  T_{00}=e^{2\frac{\rm Im(\Omega)}{r_s}ct}\left[\left\{\left(2-Z(r)\right)\left|\frac{\Omega}{r_s}\right|^2+\left(\frac{2 Q^2+r_s (r-r_s)}{2 Q^2+r r_s}\right)\frac{\Omega_0^2}{r_s^2}\right\}|T|^2|R|^2\right. \\ \left.-\frac{\left(2 Q^2+r_s (r-r_s)\right) \left(2 P^2-2 Q^2+r_s (r_s-2 r)\right)^2}{4 \left(2 P^2-r r_s\right) \left(2 Q^2+r r_s\right)^2}|T|^2|\partial_rR|^2+\frac{r_s^2 \left(r_s (r_s-r)-2 Q^2\right)}{\left(2 P^2-r r_s\right) \left(2 Q^2+r r_s\right)^2}|\partial_\theta T|^2|R|^2\right] ,       
    \end{multline}
    where
    \begin{multline}
        Z(r)=\frac{\left(2 Q^2+r_s (r-r_s)\right) \left(-2 P^2+2 Q^2+r_s (2 r+r_s)\right)}{\left(2 Q^2+r r_s\right)^2 \left(2 P^2-2 Q^2+r_s (r_s-2 r)\right)^2}\times\\\left(-2 P^2 \left(2 Q^2+r_s (r+r_s)\right)+4 Q^4+2 Q^2 r_s (3 r-2 r_s)+r_s^2 \left(2 r^2-r r_s+r_s^2\right)\right),
    \end{multline}
\end{widetext}
which is not a trivial zero. 

So, we have shown that $T_{00}$ depends on $e^{2 \frac{{\rm Im}(\Omega)}{r_s} ct}$, which diverges as $t\to\infty$ for ${\rm Im}(\Omega)>0$. 

In general, using the fact that for growing modes,
\begin{gather}
    \psi \sim e^{-i \frac{{\rm Re}(\Omega)}{r_s} ct}e^{\frac{{\rm Im}(\Omega)}{r_s} ct},\\
    \psi^* \sim e^{i \frac{{\rm Re}(\Omega)}{r_s} ct}e^{\frac{{\rm Im}(\Omega)}{r_s} ct},
\end{gather}
it is obvious that multiplication between $\partial \psi$ and $\partial \psi^*$ does not cancel the $e^{\frac{{\rm Im}(\Omega)}{r_s} ct}$ term. Therefore, every component of the stress-energy tensor diverges when ${\rm Im}(\Omega)>0$. The backreaction via Einstein equation will cause the spacetime curvature to diverge. The spacetime where CTC is allowed will be deformed and in short period of time be destroyed by the backreaction of the growing modes. Note that the charge density of the scalar field $\rho\sim |\psi|^2$ in \eqref{grofield} grows exponentially in time. The exponential particle production will inevitably deform the spacetime and we could expect some sort of phase transition of the spacetime region with CTCs. 

\section{Conclusions and Discussions}
In this work, we consider the massive and massless scalar field dynamics in the region behind the horizon of an {\it extremal} DKSBH. We begin with a short discussion on the existence of the CTCs in the extremal DKSBH spacetime followed by presenting detailed derivation and novel exact solutions to the Klein-Gordon equation in the extremal charged rotating black hole spacetime.  We find that the normal form of the radial Klein-Gordon equation completely matches the normal form of the Double Confluent Heun equation. This enables us to express the exact radial solutions in terms of the Double Confluent Heun function. 

The polynomial condition of the Double Confluent Heun function quantizes the energy of the scalar field in the extremal black hole spacetime, resulting in discrete quasiresonance frequencies. We calculate, analyze, and plot the quasiresonance frequencies of both massive and massless scalar fields. The quasiresonance frequencies are double branched, positive and negative pure imaginary, showing no propagating modes, disallow time travel, implying no violation of Hawking's CPC. The positive branch with $0\leq\Omega_0<2(n+1)$ and negative branch with $\Omega_0>2(n+1)$, which are growing exponentially, have the potential to backreact, deform and destroy spacetime region with CTCs. An analytic proof that the quasiresonances of the extreme DKSBH are pure imaginary is presented for the scalar mass less than the mass scale $M_{p}^{2}/M$. Numerical consideration for $\Omega_{0}\geq 2$ also confirm that all quasiresonances are pure imaginary, there is no propagating modes inside the extreme DKSBH where CTC exists.  

For positive $\operatorname{Im}(\Omega)$ modes, scalar fields entering the region with CTCs cause an exponential backreaction, which deforms the extremal DKS space-time and potentially destroy the original spacetime, which could lead to the truncation of geodesic, the (inner)~singularity. However, since CTC region (with positive $g_{rr}$) always exists behind the extremal horizon, the potentially produced singularity from the backreaction should be hidden behind the horizon. Even the backreaction deforms the spacetime, singularity is still not naked and (weak) cosmic cencorship could still be valid.

\begin{acknowledgments}
TB and PB are supported in part by National Research Council of Thailand~(NRCT) and Chulalongkorn University under Grant N42A660500. This research has received partial partner CaRe Global Network Project funding support from the NSRF via the Program Management Unit for Human Resources and Institutional Development, Research and Innovation [grant number B41G680025]. 
\end{acknowledgments}

\appendix
\section{Normal Form} \label{AppendixA}
The so-called Normal Form of an ordinary differential equation is the form in which an ordinary differential equation is explicitly solved for its maximum derivative. One can begin with the general form of the linear second-order ordinary differential equation as follows,
\begin{equation}
    \frac{d^2y}{dx^2}+p(x)\frac{dy}{dx}+q(x)y=0. \label{general ODE}
\end{equation}

We perform a homotopic transformation through a specific substitution for $y(x)$ that is designed to eliminate the first-order derivative term \cite{2420}, as follows,
\begin{gather}
    y=Y(x)e^{-\frac{1}{2}\int{p(x)}dx}, \label{yori}\\
    \frac{dy}{dx}=\frac{dY}{dx}e^{-\frac{1}{2}\int{p(x)}dx}-\frac{1}{2}Ype^{-\int{p(x)}dx},\\
    \frac{d^2y}{dx^2}=\frac{d^2Y}{dx^2}e^{-\frac{1}{2}\int{p(x)}dx}-\frac{1}{2}\frac{dY}{dx} pe^{-\frac{1}{2}\int{p(x)}dx}\nonumber\\-\frac{1}{2}Y\frac{dp}{dx}e^{-\frac{1}{2}\int{p(x)}dx}+\frac{1}{4}Yp^2e^{-\frac{1}{2}\int{p(x)}dx}.
\end{gather}

Substituting the expressions into \eqref{general ODE}, we obtain the following normal form,
\begin{gather}
    \frac{d^2Y}{dx^2}+\left(-\frac{1}{2}\frac{dp}{dx}-\frac{1}{4}p^2+q\right)Y=0, \label{normalform}
    \end{gather}
and once the solutions to the normal form are obtained, the solutions to the original equation \eqref{general ODE} are obtained by substituting $Y(x)$ into \eqref{yori}.

\section{The Double Confluent Heun Equation}\label{AppendixB}
Very recently, exact radial solutions to the Klein-Gordon equation in the extremal Reissner-Nordstr\"om black hole spacetime is discovered and presented in the terms of the Double Confluent Heun functions \cite{Senjaya:2024rse}. The Double Confluent Heun differential equation is a linear second-order ordinary differential equation with the following canonical form \cite{Ish,Heun},
\begin{align}
\frac{d^2y}{dx^2}+\left(\frac{\gamma}{x^2}+\frac{\delta}{x}+\epsilon\right)\frac{dy}{dx}+\left(\frac{\alpha}{x}-\frac{\beta}{x^2}\right)y &=0. \label{DCHeq}
\end{align}

The differential equation has irregular singular points at $x=0$ and $x=\infty$. Two independent solutions of this equation are given by the Double Confluent Heun functions,
\begin{align}
y &=A\operatorname{HeunD}\left(\beta,\alpha,\gamma ,\delta,\epsilon,x\right) + Be^{\frac{\gamma}{x}-\epsilon x}x^{2-\delta}\times \nonumber \\ &\phantom{=}\operatorname{HeunD}\left(-2+\beta+\delta,\alpha-2\epsilon,-\gamma ,4-\delta,-\epsilon,x\right),
\end{align}
where $\operatorname{HeunD}$ stands for the Double Confluent Heun function. 

One can transform the canonical form of the Double Confluent Heun differential equation into its normal form (by following Appendix \ref{AppendixA}) by recognizing,
\begin{gather}
p = \frac{\gamma}{x^2}+\frac{\delta}{x}+\epsilon \ \ \ , \ \ \
q = \frac{\alpha}{x}-\frac{\beta}{x^2},\\
y = Y(x)e^{-\frac{1}{2}\left(\epsilon x-\frac{\gamma}{x}\right)}x^{-\frac{\delta}{2}} \equiv \mathcal{H}(x),
\end{gather}
and obtain the normal form of \eqref{DCHeq} as follows,
\begin{align}
\frac{d^2Y}{dx^2}+K(x)Y = 0,
\end{align}
where,
\begin{align}
K(x) &= -\frac{1}{2}\frac{d{p}}{dx}-\frac{1}{4}{p}^2+{q}, \nonumber \\
&= -\frac{\epsilon^2}{4}+\frac{1}{x}\left(-\frac{\epsilon \delta}{2}+\alpha\right) + \frac{1}{x^2}\left(\frac{\delta}{2}-\frac{\delta^2+2\epsilon \gamma}{4}-\beta\right) \nonumber \\
&~~~~~+\frac{1}{x^3}\left(\gamma-\frac{\delta \gamma}{2}\right)+\frac{1}{x^4}\left(-\frac{\gamma^2}{4}\right), \label{normalDCHeq}
\end{align}
and,
\begin{multline}
    Y=e^{\frac{1}{2}\left(\epsilon x-\frac{\gamma}{x}\right)}x^{\frac{\delta}{2}} \times \\ \left[A\operatorname{HeunD}\left(\beta,\alpha,\gamma ,\delta,\epsilon,x\right)+Be^{\frac{\gamma}{x}-\epsilon x}x^{2-\delta}\times \right. \\ \left.\operatorname{HeunD}\left(-2+\beta+\delta,\alpha-2\epsilon,-\gamma ,4-\delta,-\epsilon,x\right)\right].
\end{multline}

The double confluent Heun function becomes polynomial when the following condition is met~\cite{Ish},
\begin{align}
\frac{\alpha}{\epsilon} &= -n_r,\quad n_r=0, 1,2,\dots. \label{HeunDPol}   
\end{align}

\section{Proof on The Purely Imaginary Quasiresonance}\label{AppendixC}
Let us consider the exact formula of the massive scalar quasiresonance frequency in the extremal DKSBH \eqref{QNM}. In general, $\Omega$ is complex and can be expressed as $\Omega=z=x+iy$. Substituting the expression into the left-hand side $(l.h.s.)$ of \eqref{QNM} results in this following expression,
\begin{equation}
    l.h.s= F_\pm(\Omega_0,x,y) + i G_\pm(\Omega_0,x,y), 
\end{equation}
where,
\begin{gather}
    F_\pm=\mp y+\frac{\left(2x^2-2y^2-\Omega_0^2\right) \cos\Theta+4xy \sin\Theta}{2\left(x^4 + 2 x^2 \left(y^2 - \Omega_0^2\right) + (y^2 + \
\Omega_0^2)^2\right)^{1/4}},\\ \label{ffunction}
G_\pm=\pm x+\frac{\left(-2x^2+2y^2+\Omega_0^2\right) \sin\Theta+4xy \cos\Theta}{2\left(x^4 + 2 x^2 \left(y^2 - \Omega_0^2\right) + (y^2 + \
\Omega_0^2)^2\right)^{1/4}}, \\ \label{gfunction}
\Theta=\frac{1}{2}\operatorname{Arg}\left[\Omega_0^2-(x + i y)^2 \right].
\end{gather}

The right-hand side of \eqref{QNM} is a negative integer, which is strictly real. This implies the imaginary component of the $l.h.s.$ has to be equal to zero,
\begin{equation}
G_\pm=0,  \label{im} 
\end{equation}
and leaving,
\begin{equation}
F_\pm=-(n+1)\leq -1.  \label{real} 
\end{equation}
Therefore, one obvious solution is $x=0, \Omega = iy$, pure imaginary mode.  

Remark that the denominator of the second terms can be rearranged as the following,
\begin{multline}
D_{FG}=x^4 + 2 x^2 \left(y^2 - \Omega_0^2\right) + (y^2 + \Omega_0^2)^2 \\= (\Omega_0^2-x^2)^2+y^2(2x^2+2\Omega_0^2+y^2)>0,
\end{multline}
and it is obvious that $D_{FG}$ is positive definite.

Now, let us consider $\Theta$,
\begin{gather}
  \Theta=\frac{1}{2} \operatorname{Arg}\left[\Omega_0^2-x^2+y^2-2ixy\right]=-\frac{1}{2}\tan^{-1} \chi,  \\ \label{theta1}
  \Theta^*=\frac{1}{2} \operatorname{Arg}\left[\Omega_0^2-x^2+y^2+2ixy\right]=\frac{1}{2}\tan^{-1} \chi,\\ \label{theta2}
  \chi=\frac{2xy}{\Omega_0^2-x^2+y^2}, 
\end{gather}
where $\chi$ is real and additionally, we also have these following relations,
\begin{gather}
    \Theta^*=-\Theta,\\ \label{theta3}
    \Theta(-x)=-\Theta(x).
\end{gather}

Since $x$ has domain as the following,
\begin{equation}
    0\leq x<\Omega_0, \label{xrange}
\end{equation}
with $y \ \operatorname{\epsilon} \ \mathcal{R}$, we find $\chi \  \operatorname{\epsilon} \ \mathcal{R}$. However, $\operatorname{tan}^{-1} \chi$ has value ranged between $\left(-\frac{\pi}{2},\frac{\pi}{2}\right)$ and together with the factor $\frac{1}{2}$, $\Theta$ is ranged between $\left(-\frac{\pi}{4},\frac{\pi}{4}\right)$, therefore, we find,
\begin{gather}
    \frac{1}{\sqrt{2}}>\cos\Theta \geq 1, \label{cos}\\ 
    {\sin  \Theta \ }=\left\{ \begin{array}{c}
<0\ \ \ ,\ \ \ y>0 \\ 
>0\ \ \ ,\ \ \ y<0 \end{array}
\right. . \label{sin}
\end{gather}

Start with equation \eqref{real}, it requires $F_\pm$ to be less than or equal to $-1$. This condition will also be a constraint for $\{x,y\}$ in each of positive and negative modes. To investigate this, let us explore the derivatives of $F_\pm$ as follows,
\begin{widetext}
 \begin{multline}
 \partial_x F_\pm=\frac{D_{FG}^{-5/4}}{2}\left[x\left\{(3\Omega_0^2-2x^2)(\Omega_0^2-x^2)+3y^2\Omega_0^2+2y^2(y^2+2x^2)\right\} \cos  \Theta  \right. \\ \left. +y \left\{(3\Omega_0^2-2x^2)(\Omega_0^2-x^2)+\Omega_0^2(2x^2+5y^2)+2y^2(y^2+2x^2)\right\} \sin\Theta\right],    \label{dxF}
 \end{multline} 
 \begin{equation}
 \partial^2_{x}F_\pm=\frac{3D_{FG}^{-9/4}}{2}\left[\left\{x^4 + (\Omega_0^2 + y^2)^2 - 2 x^2 (\Omega_0^2 + 3 y^2)\right\} \cos  \Theta  +4 x y (\Omega_0^2-x^2+y^2) \sin\Theta\right],      \label{d2xF}
 \end{equation} 
  \begin{multline}
 \partial_y F_\pm=\mp 1+\frac{D_{FG}^{-5/4}}{2}\left[x\left\{(3\Omega_0^2-2x^2)(\Omega_0^2-x^2)+3y^2\Omega_0^2+2y^2(y^2+2x^2)\right\} \sin  \Theta  \right. \\ \left. -y \left\{(3\Omega_0^2-2x^2)(\Omega_0^2-x^2)+\Omega_0^2(2x^2+5y^2)+2y^2(y^2+2x^2)\right\} \cos\Theta\right],  \label{dyF}
 \end{multline} 
  \begin{equation}
 \partial^2_{y}F_\pm=-\frac{3D_{FG}^{-9/4}}{2}\left[\left\{x^4 + (\Omega_0^2 + y^2)^2 - 2 x^2 (\Omega_0^2 + 3 y^2)\right\} \cos  \Theta  +4 x y (\Omega_0^2-x^2+y^2) \sin\Theta\right].  \label{d2yF}    
 \end{equation} 
\end{widetext}

Let us consider $\partial_x F_\pm$ and $\partial^2_x F_\pm$. It is obvious that the terms in the curly brackets of \eqref{dxF} is definite positive and we can collect this following information,
\begin{itemize}
    \item The first term inside the square bracket \eqref{dxF} is positive for $x>0$ and zero for $x=0$.
    \item The second term inside the square bracket \eqref{dxF} is an even function of $y$ and always negative except for $x=0$, where it gives zero.   
    \item Along fixed $y$, $F_\pm (x)$ starts off with zero slope at $x=0$,
    \begin{gather}
      \partial_{x}F_\pm (x=0)=0,\\
      F_\pm (x=0)=\mp y-\frac{2y^2+\Omega_0^2}{2(y^2+\Omega_0^2)^{1/2}},
    \end{gather}
    and as $x\to+\infty$, $F_\pm (x)$ continuously increases and becomes asymptotically,
    \begin{gather}
    \partial_x F_\pm (x=+\infty)=1,\\
    F_{\pm}(x=+\infty)=\infty.
    \end{gather}
\end{itemize}

Based on the above analysis, we can conclude that there is a minimum at $x=0$ for every fixed $y$. The curvature at that point reads,
    \begin{equation}
        \partial^2_{x}F_\pm (x=0)=\frac{3\Omega_0^4}{2(\Omega_0^2+y^2)^{5/2}}>0,
    \end{equation}
allowing us to estimate the shape of $F_\pm$ for any constant $y$, i.e., concave upward with respect to $x$. 
   
Now, let us move on to $\partial_y F_\pm$ and $\partial^2_y F_\pm$,
\begin{itemize}
    \item The first term inside the square bracket \eqref{dyF} is positive (negative) for $y<0~(y>0)$.
    \item The second term inside the square bracket \eqref{dyF} is an odd function of $y$ that is positive (negative) for $y > 0~(y<0)$.
    \item Along constant $x$, the positive mode $F_+(y)$ begins with zero and zero slope at $y=-\infty$. It reaches $y=0$ with slope $\partial_y F_+=-1$ and continuously decreases with saturated slope, $\partial_y F_+=-2$, at $y=+\infty$, where $F_+=-\infty$.
   \item The negative mode, $F_-(y)$, starts with $F_-=-\infty$ at $y=-\infty$ with saturated slope $\partial_y F_-=2$. It reaches $y=0$ with slope $\partial_yF_-=1$ and saturated to zero with zero slope at $y= +\infty$.
\end{itemize}



It is also straightforward to check that,
\begin{gather}
    F_{\pm}(x,0)=\frac{-\Omega_0^2+2x^2}{2 \sqrt{\Omega_0^2-x^2}}, \label{Fx0}
\end{gather}
is minimized by $x=0$,
\begin{equation}
    F_{+}(0,0)=F_{-}(0,0)=-\frac{\Omega_{0}}{2}.
\end{equation}

Based on above investigation, we can recap as follows,
\begin{itemize}
    \item The positive mode, $F_+$, is monotonically decreasing in the positive $y$ direction and tends to be flat and zero in the region $y<0$.  It fulfills \eqref{real} in the region $y>0$.
    \item The negative mode, $F_-$, is monotonically increasing in the positive $y$ direction and tends to be flat and zero in the region $y>0$.  It fulfills \eqref{real} in the region $y<0$.
    \item Therefore for $\Omega_{0}<2$ along $x=0$ direction, {\it condition $y>0~(y<0)$ is required for positive~(negative) mode} to fulfill \eqref{real} respectively. 
    \item Remark that equation \eqref{Fx0} tells us $F_\pm (y)$ is less negative for $x>0$. The positive~(negative) modes fulfill  \eqref{real} later at larger $y>0~(y<0)$. Consequently, the requirement $y>0~(y<0)$ for positive~(negative) mode holds for all $x>0$. 
\end{itemize} 

After discussing $F_\pm$, let us move on to $G_\pm$. Along a constant $y$, $G_\pm (x)$ is an odd function that is symmetric with respect to $(x \leftrightarrow -x)$. And due to the nature of an odd function, $G_\pm (x)$ is always zero at $x=0$. We will show that this is the only solution which satisfies both \eqref{im} and \eqref{real}. It can be done by demonstrating that $G_{\pm}$ has derivatives which do not change signs in the relevant region. 

To investigate the behavior of $G_\pm(x)$, let us consider its first and second derivatives with respect to $x$ and $y$ as follows,
\begin{widetext}
 \begin{multline}
 \partial_x G_\pm=\pm 1+\frac{D_{FG}^{-5/4}}{2}\left[-x\left\{(3\Omega_0^2-2x^2)(\Omega_0^2-x^2)+3y^2\Omega_0^2+2y^2(y^2+2x^2)\right\} \sin  \Theta  \right. \\ \left. +y \left\{(3\Omega_0^2-2x^2)(\Omega_0^2-x^2)+\Omega_0^2(2x^2+5y^2)+2y^2(y^2+2x^2)\right\} \cos\Theta\right],   \label{dxG}
 \end{multline} 
  \begin{equation}
      \partial^2_x G_\pm = \frac{3D_{FG}^{-9/4}}{2}\left[-\left\{x^4 + (\Omega_0^2 + y^2)^2 - 2 x^2 (\Omega_0^2 + 3 y^2)\right\} \sin  \Theta  +4 x y (\Omega_0^2-x^2+y^2) \cos\Theta\right],      \label{d2xG}
  \end{equation}
\begin{multline}
 \partial_y G_\pm=\frac{D_{FG}^{-5/4}}{2}\left[x\left\{(3\Omega_0^2-2x^2)(\Omega_0^2-x^2)+3y^2\Omega_0^2+2y^2(y^2+2x^2)\right\} \cos  \Theta  \right. \\ \left. +y \left\{(3\Omega_0^2-2x^2)(\Omega_0^2-x^2)+\Omega_0^2(2x^2+5y^2)+2y^2(y^2+2x^2)\right\} \sin\Theta\right],    \label{dyG}
 \end{multline} 
   \begin{equation}
      \partial^2_y G_\pm = \frac{3D_{FG}^{-9/4}}{2}\left[\left\{x^4 + (\Omega_0^2 + y^2)^2 - 2 x^2 (\Omega_0^2 + 3 y^2)\right\} \sin  \Theta  -4 x y (\Omega_0^2-x^2+y^2) \cos\Theta\right].      \label{d2yG}
  \end{equation}
\end{widetext}

Let us look into the equation \eqref{dyG}. The first term within the square bracket is always positive while the second term is either zero, when $y=0$, or negative otherwise. Therefore, we can estimate the behavior of $\partial_y G_\pm$ as follows,
\begin{itemize}

\item With respect to $y$,  if $x=0$, $\partial_y G_\pm$ is zero for all $y$.
\item For each non-zero $x$ in the range $0\ < \ x <\Omega_0$, $G_\pm$ starts off with a maximum positive slope at $y=0$,
\begin{gather}
\partial_y G_\pm(y=0) = \frac{x(3\Omega_0^2-2x^2)}{2(\Omega_0^2-x^2)^{3/2}}  > 0, 
\end{gather}
with,
\begin{gather}
\partial^2_y G_\pm(y=0) = 0.
\end{gather}
\item As $|y|$ grows, the negative second term in \eqref{dyG} increases, reducing the slope until $G_\pm$ becomes flat at $|y|=\infty$, where the corresponding slope is,
\begin{gather}
    \partial_y G_\pm(y=\pm\infty) =0.
\end{gather}
\end{itemize}

Namely, $G_{\pm}$ monotonically increases with respect to $y$, from zero slope at $y=-\infty, G_{\pm}= 0,-2x$, crosses $y=0$ with $G_{\pm}=\pm x$ and saturates to $G_{\pm}=2x,0$ with zero slope at $y=+\infty$ respectively. 

With respect to $x$, for $y>0$, $\partial_x G_+$ is always positive since $-x\sin\Theta >0$ and $ y\cos\Theta >0$~(note that this requires $\Omega_{0}<2$ for $F_{+}$ to satisfy \eqref{real}). Thus, $G_+(x)$ is monotonically increasing in positive $x$ direction, starts with zero at $x=0$ and continuously climbs up as $x$ increases. Conversely for the negative modes, $\partial_x G_-$ is always negative for $y<0$~(also this requires $\Omega_{0}<2$ for $F_{-}$ to satisfy \eqref{real}) because $-x\sin\Theta <0$ and $ y\cos\Theta <0$. $G_-$ starts with zero at $x=0$ and continuously drops down as $x$ increases. Therefore, along fixed $y$, the only zero of $G_\pm$ is at $x=0$.

Based on these analyses, it can be concluded that there is only one solution for $G_\pm$ that satisfies \eqref{im}, i.e., at $x=0$ for $\Omega_{0}<2$. As a result, for $\Omega_{0}<2$ the quasiresonance frequency is purely imaginary, $\Omega=\pm iy$. For $\Omega_{0}\geq 2$, we numerically explore the solutions and also find that they are strictly pure imaginary as shown in Fig.~\ref{contourfig}.


\begin{figure*}[h!]
    \centering
    \includegraphics[width=10cm]{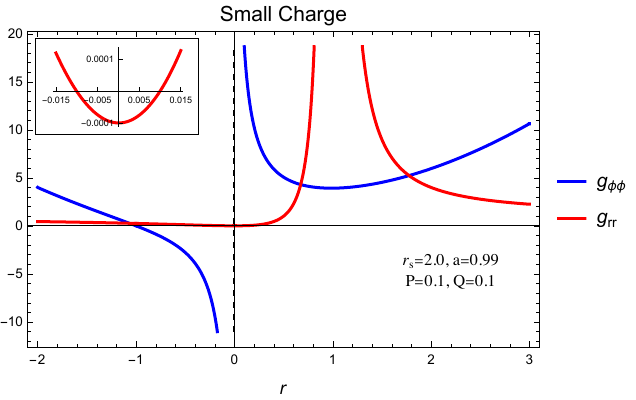}
    \caption{Behavior of $g_{rr}$ and $g_{\phi\phi}$ as functions of radial coordinate $r$ in small charge configuration.} 
  \label{fig3}
\end{figure*}

\begin{figure*}[h!]
    \centering
    \includegraphics[width=10cm]{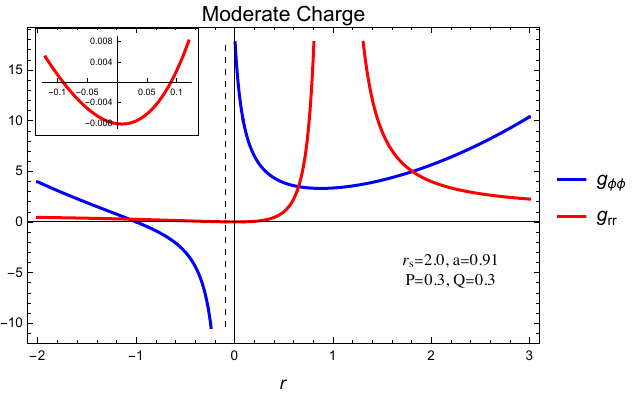}
    \caption{Behavior of $g_{rr}$ and $g_{\phi\phi}$ as functions of radial coordinate $r$ in moderate charge configuration.} 
  \label{fig4}
\end{figure*}

\begin{figure*}[h!]
    \centering
    \includegraphics[width=10cm]{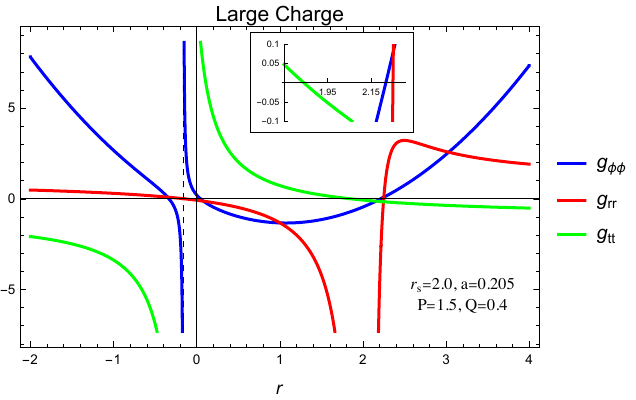}
    \caption{Behavior of $g_{rr}$ and $g_{\phi\phi}$ as functions of radial coordinate $r$ in large charge configuration.} 
  \label{fig5}
\end{figure*}

\begin{figure*}[t!]
        \subfloat[$G_{+}=0$ contour.]{%
        \includegraphics[width=.48\linewidth]{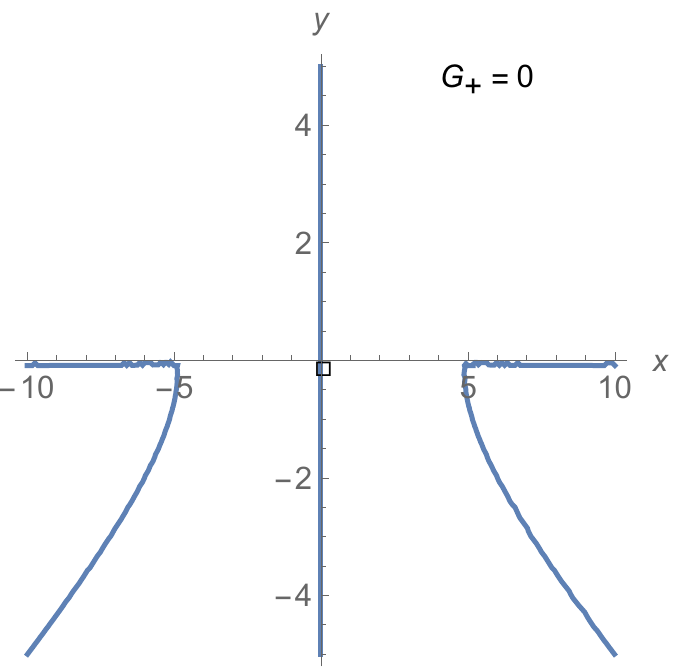}
            \label{Gpfig}
        }\hfill
        \subfloat[$G_{-}=0$ contour.]{%
        \includegraphics[width=.48\linewidth]{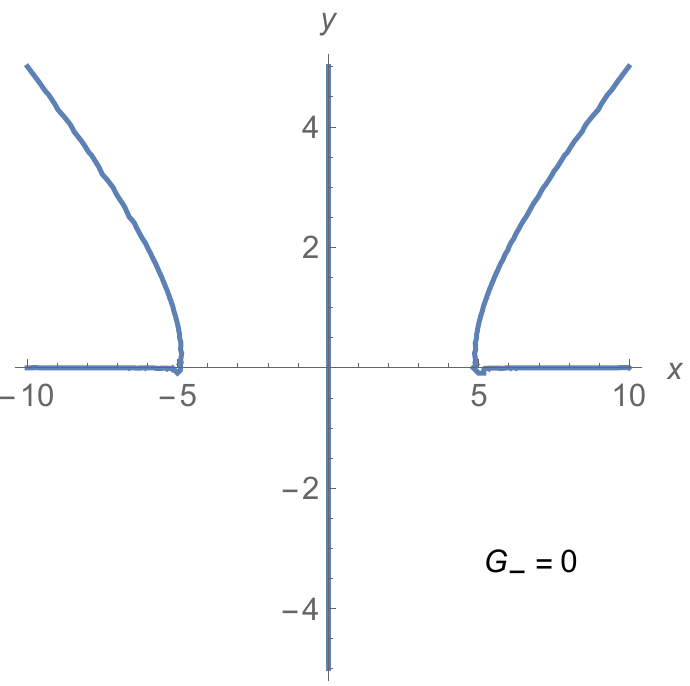}
            \label{Gmfig}
        }\\
        \subfloat[$F_{+}<-(n+1)$ for $n=0-3$.]{%
        \includegraphics[width=.48\linewidth]{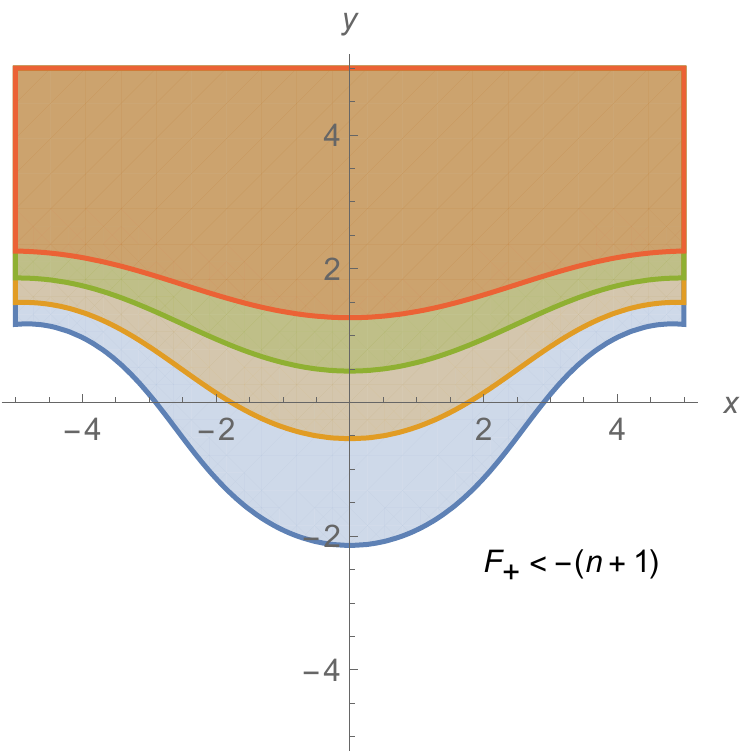}
            \label{Fpfig}
        }\hfill
        \subfloat[$F_{-}<-(n+1)$ for $n=0-3$.]{%
        \includegraphics[width=.48\linewidth]{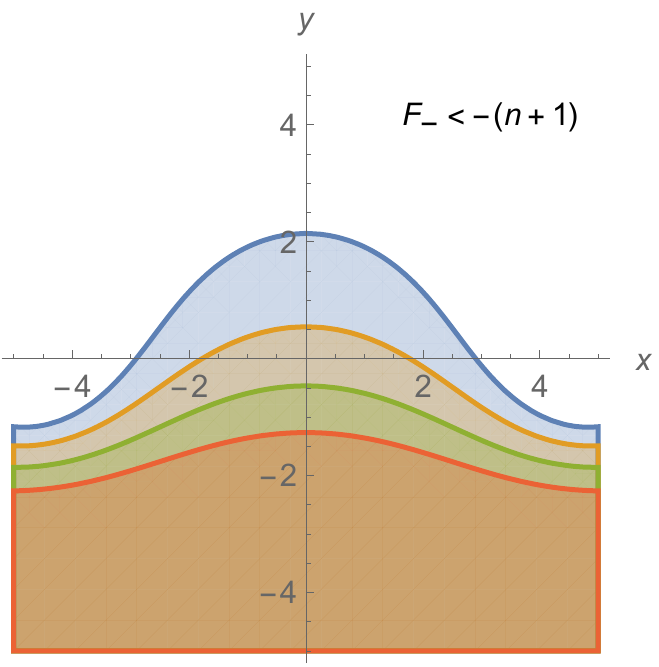}
            \label{Fmfig}
        }
       \caption{Region of quasiresonances satisfying \eqref{im} and\eqref{real} for $\Omega_{0}=5$.}
    \label{contourfig}
\end{figure*}

\begin{thebibliography}{99}
\bibitem{Hawking}
S.~W.~Hawking,
``The Chronology protection conjecture,''
Phys. Rev. D \textbf{46}, 603-611 (1992)
doi:10.1103/PhysRevD.46.603

\bibitem{Vis92}
M.~Visser,
``From wormhole to time machine: Comments on Hawking's chronology protection conjecture,''
Phys. Rev. D \textbf{47}, 554-565 (1993)
doi:10.1103/PhysRevD.47.554
[arXiv:hep-th/9202090 [hep-th]].

\bibitem{Woodward}
J.~F.~Woodward,
Found. Phys. Lett. \textbf{8}, 1-39 (1995)
doi:10.1007/BF02187529

\bibitem{Shinkai}
H.~a.~Shinkai and S.~A.~Hayward,
``Fate of the first traversible wormhole: Black hole collapse or inflationary expansion,''
Phys. Rev. D \textbf{66}, 044005 (2002)
doi:10.1103/PhysRevD.66.044005
[arXiv:gr-qc/0205041 [gr-qc]].

\bibitem{Novikov}
I.~D.~Novikov and D.~I.~Novikov,
``Collapse of a Wormhole and its Transformation into Black Holes,''
J. Exp. Theor. Phys. \textbf{129}, no.4, 495-502 (2019)
doi:10.1134/S1063776119100248

\bibitem{Ashok}
A.~Das,
Field Theory: A Path Integral Approach (2nd Edition), World Scientific Publishing Company (2006).

\bibitem{Boris}
B.~M.~Karnakov, V.~P.~Krainov
Field Theory: A Path Integral Approach (2nd Edition), World Scientific Publishing Company (2006).

\bibitem{Bunyaratavej:2024qgk}
T.~Bunyaratavej, P.~Burikham and D.~Senjaya,
``Revisiting chronology protection conjecture in the Dyonic Kerr\textendash{}Sen black hole spacetime,''
Eur. Phys. J. C \textbf{85} (2025) no.3, 270
doi:10.1140/epjc/s10052-025-13935-2
[arXiv:2408.06023 [gr-qc]].

\bibitem{Richartz}
M.~Richartz,
``Quasinormal modes of extremal black holes,''
Phys. Rev. D \textbf{93}, no.6, 064062 (2016)
doi:10.1103/PhysRevD.93.064062
[arXiv:1509.04260 [gr-qc]].

\bibitem{Joykutty}
J.~Joykutty,
``Existence of Zero-Damped Quasinormal Frequencies for Nearly Extremal Black Holes,''
Annales Henri Poincare \textbf{23}, no.12, 4343-4390 (2022)
doi:10.1007/s00023-022-01202-z
[arXiv:2112.05669 [gr-qc]].

\bibitem{NE1}
S.~Hod,
``Quasi-bound state resonances of charged massive scalar fields in the near-extremal Reissner\textendash{}Nordstr\"om black-hole spacetime,''
Eur. Phys. J. C \textbf{77}, no.5, 351 (2017)
doi:10.1140/epjc/s10052-017-4920-8
[arXiv:1705.04726 [hep-th]].

\bibitem{NE2}
S.~Hod,
``Numerical evidence for universality in the relaxation dynamics of near-extremal Kerr\textendash{}Newman black holes,''
Eur. Phys. J. C \textbf{75}, no.12, 611 (2015)
doi:10.1140/epjc/s10052-015-3845-3
[arXiv:1511.05696 [hep-th]].

\bibitem{NE3}
P.~Burikham, S.~Ponglertsakul and T.~Wuthicharn,
``Quasi-normal modes of near-extremal black holes in generalized spherically symmetric spacetime and strong cosmic censorship conjecture,''
Eur. Phys. J. C \textbf{80}, no.10, 954 (2020)
doi:10.1140/epjc/s10052-020-08528-0
[arXiv:2010.05879 [gr-qc]].

\bibitem{Ponglertsakul:2020ufm}
S.~Ponglertsakul and B.~Gwak,
``Massive scalar perturbations on Myers-Perry\textendash{}de Sitter black holes with a single rotation,''
Eur. Phys. J. C \textbf{80}, no.11, 1023 (2020)
doi:10.1140/epjc/s10052-020-08616-1
[arXiv:2007.16108 [gr-qc]].

\bibitem{Wu}
D.~Wu, S.~Q.~Wu, P.~Wu and H.~Yu,
``Aspects of the dyonic Kerr-Sen- AdS$_4$ black hole and its ultraspinning version,''
Phys. Rev. D \textbf{103} (2021) no.4, 044014
doi:10.1103/PhysRevD.103.044014
[arXiv:2010.13518 [gr-qc]].

\bibitem{Jana}
S.~Jana and S.~Kar,
``Shadows in dyonic Kerr-Sen black holes,''
Phys. Rev. D \textbf{108} (2023) no.4, 044008
doi:10.1103/PhysRevD.108.044008
[arXiv:2303.14513 [gr-qc]].

\bibitem{Sakti:2022izj}
M.~F.~A.~R.~Sakti and P.~Burikham,
Phys. Rev. D \textbf{106} (2022) no.10, 106006
doi:10.1103/PhysRevD.106.106006
[arXiv:2206.10868 [hep-th]].

\bibitem{Senjaya}
D.~Senjaya, P.~Burikham and T.~Harko,
``The exact relativistic scalar quasibound states of the dyonic Kerr\textendash{}Sen black hole: quantized energy, and Hawking radiation,''
Eur. Phys. J. C \textbf{84}, no.8, 857 (2024)
doi:10.1140/epjc/s10052-024-13225-3
[arXiv:2405.15219 [gr-qc]].

\bibitem{Chandrasekhar}
S.~Chandrasekhar,
``The mathematical theory of black holes,'' Clarendon Press, 1985, ISBN 978-019-85-0370-5

\bibitem{Derig}
A.~A.~Deriglazov and B.~F.~Rizzuti,
``Classical mechanics in reparametrization-invariant formulation and the Schr\"odinger equation,''
Am. J. Phys. \textbf{79}, 882-885 (2011)
doi:10.1119/1.3593270
[arXiv:1105.0313 [math-ph]].

\bibitem{Luzio}
J.~L.~Lucio-M, J.~A.~Nieto, J.~D.~Vergara,
``A generalized Klein-Gordon equation from a reparametrized Lagrangian,``
Phys. Lett. A. \textbf{219} (1996), 150-154
doi:0375-9601(96)00456-2

\bibitem{Dong}
S.~H.~Dong,
``Wave equations in higher dimensions,''
Springer, 2011,
ISBN 978-94-007-1916-3
doi:10.1007/978-94-007-1917-0

\bibitem{35}
W.~W.~Bell, Special Functions for Scientists and Engineers, Courier Corporation (2004).

\bibitem{Press}
W.~H.~Press and S.~A.~Teukolsky,
``Perturbations of a Rotating Black Hole. II. Dynamical Stability of the Kerr Metric,''
Astrophys. J. \textbf{185}, 649-674 (1973)
doi:10.1086/152445

\bibitem{Berti1}
E.~Berti, V.~Cardoso and M.~Casals,
``Eigenvalues and eigenfunctions of spin-weighted spheroidal harmonics in four and higher dimensions,''
Phys. Rev. D \textbf{73}, 024013 (2006)
[erratum: Phys. Rev. D \textbf{73}, 109902 (2006)]
doi:10.1103/PhysRevD.73.109902
[arXiv:gr-qc/0511111 [gr-qc]].

\bibitem{Berti2}
E.~Berti, V.~Cardoso and M.~Casals,
``Eigenvalues and eigenfunctions of spin-weighted spheroidal harmonics in four and higher dimensions,''
Phys. Rev. D \textbf{73}, 024013 (2006)
[erratum: Phys. Rev. D \textbf{73}, 109902 (2006)]
doi:10.1103/PhysRevD.73.109902
[arXiv:gr-qc/0511111 [gr-qc]].

\bibitem{Cho:2009wf}
H.~T.~Cho, A.~S.~Cornell, J.~Doukas and W.~Naylor,
``Asymptotic iteration method for spheroidal harmonics of higher-dimensional Kerr-(A)dS black holes,''
Phys. Rev. D \textbf{80}, 064022 (2009)
doi:10.1103/PhysRevD.80.064022
[arXiv:0904.1867 [gr-qc]].

\bibitem{Suzuki:1998vy}
H.~Suzuki, E.~Takasugi and H.~Umetsu,
``Perturbations of Kerr-de Sitter black hole and Heun's equations,''
Prog. Theor. Phys. \textbf{100}, 491-505 (1998)
doi:10.1143/PTP.100.491
[arXiv:gr-qc/9805064 [gr-qc]].

\cite{Baumgarte:2010ndz}
\bibitem{Baumgarte:2010ndz}
T.~W.~Baumgarte and S.~L.~Shapiro,
Cambridge University Press, 2010,
doi:10.1017/CBO9781139193344


\bibitem{2420}
G.~F.~Simmons,
Differential Equations with Applications and Historical Notes, CRC Press (2016).

\bibitem{Heun}
A.~Ronveaux,
Heun's Differential Equations, Clarendon Press (1995).

\bibitem{Senjaya:2024rse}
D.~Senjaya and S.~Ponglertsakul,
``The extreme Reissner\textendash{}Nordstr\"om Black Hole: New exact solutions to the Klein\textendash{}Gordon equation with minimal coupling,''
Annals Phys. \textbf{473}, 169898 (2025)
doi:10.1016/j.aop.2024.169898
[arXiv:2405.07579 [gr-qc]].

\bibitem{Ish}
T.~A.~Ishkhanyan, V.~A.~Manukyan, A.~H.~Harutyunyan and A.~M.~Ishkhanyan,
''Confluent hypergeometric expansions of the solutions of the double-confluent Heun equation,''
Armenian J. Phys. \textbf{10}, 212-223 (2017)
\end{thebibliography}
\end{document}